\documentclass[sigconf, screen]{acmart}
\pdfoutput=1
\usepackage[utf8]{inputenc}
\usepackage{color}

\usepackage{centernot}

\usepackage{siunitx}
\usepackage{textcomp}
\usepackage{caption}

\usepackage{pst-func}
\usepackage{amssymb}
\usepackage{graphicx}

\usepackage{paralist}
\usepackage{environ}
\usepackage{pgfplots}
\usepackage{tikz}

\usepackage{mathtools}
\usepackage{array,multirow}

\usepackage{url}

\usepackage{pgfplotstable}
\usepackage{threeparttable}

\usepackage{etoolbox}
\usepackage{arydshln}

\usepackage{xspace} 
 
\usepackage{dirtytalk} 
\usepackage{stmaryrd} 

\usepackage[inline,shortlabels]{enumitem} 
\newlist{inlinelist}{enumerate*}{1}
\setlist*[inlinelist,1]{%
  label=(\roman*),
}

\usepackage{xifthen}  
\usepackage{ifthen}

\usepackage{nicefrac}

\usepackage{hyperref}
\usepackage{cleveref}
\usepackage{bigints}

\usetikzlibrary{shadows}
\usetikzlibrary{pgfplots.dateplot}
\usetikzlibrary{matrix,chains,positioning,decorations.pathreplacing}
\usetikzlibrary{patterns,calc,fit,arrows}
\usetikzlibrary{shapes.geometric,shapes.misc}

\tikzset{
  block/.style={
  draw, 
  rectangle, 
  minimum height=1.5cm, 
  minimum width=2.5cm, 
  align=center,
  draw=red!50!black!50, 
  top color=white, 
  bottom color=red!50!black!20, 
  }, 
  line/.style={->,>=latex'}
}

\tikzset{
  inputblock/.style={
  draw, 
  rounded rectangle, 
  minimum height=1.2cm, 
  minimum width=2.5cm, 
  align=center,
  very thick,draw=black!50,
  top color=white,bottom color=black!20,
  }, 
  line/.style={->,>=latex'}
}

\tikzset{
  outputblock/.style={
  draw, 
  rounded rectangle, 
  minimum height=1.2cm, 
  minimum width=2.5cm, 
  align=center,
  very thick,draw=black!50,
  top color=white,bottom color=blue!20,
  }, 
  line/.style={->,>=latex'}
}

\usepackage[final,nomargin,inline,index]{fixme} 
\fxusetheme{color}

\FXRegisterAuthor{bart}{anbart}{\color{magenta} {\underline{bart}}}
\FXRegisterAuthor{tizi}{antizi}{\color{red} {\underline{tizi}}}
\FXRegisterAuthor{ste}{anste}{\color{blue} {\underline{ste}}}
\FXRegisterAuthor{nico}{annico}{\color{ForestGreen} {\underline{nico}}}

\hypersetup{
  breaklinks   = true,
  colorlinks   = true, 
  urlcolor     = blue, 
  linkcolor    = blue, 
  citecolor    = red   
}
\hypersetup{final} 

\usepackage{listings}

\definecolor{listingBG}{HTML}{FFFFCB}%
\definecolor{listingFrame}{HTML}{BBBB98}%
\definecolor{listingLineno}{rgb}{0.5,0.5,1.0}%

\definecolor{LightGrey}{rgb}{0.975,0.975,0.975}

\lstdefinelanguage{bitml}{
	commentstyle=\color{Gray},
	morecomment=[l]{;},
	morecomment=[s]{/*}{*/},
	classoffset=0,
        escapechar=\$,
	morekeywords={contract,pre,choice,reveal,split,put,revealif,withdraw,after},
	keywordstyle=\color{Blue}\bfseries,
	classoffset=1,
	morekeywords={deposit,secret,participant,sig,versig,fun,unit,int,string,bool,address,uint},
	keywordstyle=\color{TealBlue},
	classoffset=2,
	morekeywords={BTC,true},
	keywordstyle=\color{Plum}\bfseries,
}

\lstdefinelanguage{solidity}{
	commentstyle=\color{Gray},
	morecomment=[l]{//},
	morecomment=[s]{/*}{*/},
	classoffset=0,
        escapechar=\$,
	morekeywords={struct,mapping,function,this,public,private,static,final,class,extends,switch,case,break,finally,try,catch,return,if,else,new},
	keywordstyle=\color{Blue}\bfseries,
	classoffset=1,
	morekeywords={unit,int,string,bool,address,uint},
	keywordstyle=\color{TealBlue},
	classoffset=2,
	morekeywords={ether,wei,finney,contract,send,throw,msg,sender,value},
	keywordstyle=\color{Plum}\bfseries,
}

\lstdefinelanguage{java}{
	escapechar=\$,
        commentstyle=\color{Gray},
	morecomment=[l]{//},
	morecomment=[s]{/*}{*/},
	morestring=[b]",
        classoffset=0,
	morekeywords={public,private,static,final,class,extends,switch,case,break,finally,try,catch,void,int,boolean,throws,throw,return,if,else,new},
	keywordstyle=\color{keyword}\bfseries
}

\newcommand{\ifempty}[3]{%
  \ifthenelse{\isempty{#1}}{#2}{#3}%
}

\newcommand{\ifdots}[3]{%
  \ifthenelse{\equal{#1}{...}}{#2}{#3}%
}

\newcommand{\hidden}[1]{}

\definecolor{BlueViolet}{rgb}{0, 0, 0.55}
\definecolor{RubineRed}{rgb}{0.88, 0.07, 0.37}
\definecolor{ForestGreen}{rgb}{0.13, 0.55, 0.13}
\definecolor{NavyBlue}{rgb}{0.0, 0.0, 0.52}
\definecolor{Black}{rgb}{0.02, 0.02, 0.02}
\definecolor{TealBlue}{rgb}{0, 0.52, 0.52}
\definecolor{Blue}{rgb}{0, 0, 1}
\definecolor{MidnightBlue}{rgb}{0.0, 0.2, 0.4}
\definecolor{Gray}{rgb}{0.41, 0.41, 0.41}

\newcommand{\Real}[1]{\mathrm{Real}}

\newcommand{\codefont}{\fontsize{9}{9}\selectfont}
\newcommand{\code}[1]{{\tt\codefont {#1}}}

\newcommand{\lineno}[1]{{\tt\codefont {\textcolor{NavyBlue}{#1}}}}

\newcommand{\codebackslash}{\symbol{`\\}}

\def\etc{etc.\@\xspace}
\newcommand{\Eg}{E.g.\@\xspace}
\newcommand{\eg}{e.g.\@\xspace}
\newcommand{\ie}{i.e.\@\xspace}

\theoremstyle{plain}

\theoremstyle{definition}

  {}

\newcommand{\BTC}{\textup{%
  \leavevmode
  \vtop{\offinterlineskip 
    \setbox0=\hbox{B}%
    \setbox2=\hbox to\wd0{\hfil\hskip-.03em
    \vrule height .3ex width .15ex\hskip .08em
    \vrule height .3ex width .15ex\hfil}
    \vbox{\copy2\box0}\box2}}\xspace}

\DeclareMathAlphabet{\mathbfsf}{\encodingdefault}{\sfdefault}{bx}{n}

\crefname{appendix}{appendix}{appendices}
\Crefname{appendix}{Appendix}{Appendices}
\crefname{notation}{notation}{notations}
\Crefname{notation}{Notation}{Notations}

\definecolor{LightGrey}{rgb}{0.95,0.95,0.95}
\definecolor{keyword}{HTML}{7F0055}

\newlength\replength
\newcommand\repfrac{.1}

\newcommand\rulewidth{.6pt}
\newcommand\tdashfill[1][\repfrac]{\cleaders\hbox to \replength{%
  \smash{\rule[\arraystretch\ht\strutbox]{\repfrac\replength}{\rulewidth}}}\hfill}

\newcommand\tdotfill[1][\repfrac]{\cleaders\hbox to \replength{%
  \smash{\raisebox{\arraystretch\dimexpr\ht\strutbox-.1ex\relax}{.}}}\hfill}

\newcommand{\contrAdvC}[2]{\mathcal{C}} 

\usepackage{anyfontsize}
\AtBeginEnvironment{thebibliography}{\balance}

\setcopyright{acmcopyright}
\acmPrice{15.00}
\acmDOI{10.1145/3338906.3341173}
\acmYear{2019}
\copyrightyear{2019}
\acmISBN{978-1-4503-5572-8/19/08}
\acmConference[ESEC/FSE '19]{Proceedings of the 27th ACM Joint European Software Engineering Conference and Symposium on the Foundations of Software Engineering}{August 26--30, 2019}{Tallinn, Estonia}
\acmBooktitle{Proceedings of the 27th ACM Joint European Software Engineering Conference and Symposium on the Foundations of Software Engineering (ESEC/FSE '19), August 26--30, 2019, Tallinn, Estonia}

\begin{document}

\title{Developing Secure Bitcoin Contracts with BitML}

\author[N. Atzei]{Nicola Atzei}
\affiliation{
	\institution{University of Cagliari} \country{Italy}
	}
\email{atzeinicola@gmail.com}

\author[M. Bartoletti]{Massimo Bartoletti}
\affiliation{\institution{University of Cagliari} \country{Italy}}
\email{bart@unica.it}

\author[S. Lande]{Stefano Lande}
\affiliation{\institution{University of Cagliari} \country{Italy}}
\email{lande@unica.it}

\author[N. Yoshida]{Nobuko Yoshida}
\affiliation{\institution{Imperial College London} \country{UK}}
\email{n.yoshida@imperial.ac.uk}

\author[R. Zunino]{Roberto Zunino}
\affiliation{\institution{University of Trento} \country{Italy}}
\email{roberto.zunino@unitn.it}

\begin{abstract}
  We present a toolchain for developing and verifying smart contracts
  that can be executed on Bitcoin.
  The toolchain is based on BitML, a recent domain-specific language 
  for smart contracts with a computationally sound embedding into Bitcoin.
  Our toolchain automatically verifies relevant properties of contracts,
  among which \emph{liquidity}, 
  ensuring that funds do not remain frozen within a contract forever.
  A compiler is provided to translate BitML contracts
  into sets of standard Bitcoin transactions:
  executing a contract corresponds to appending these transactions to the blockchain.
  We assess our toolchain through a benchmark of representative contracts.

  \begin{flushright}
    \emph{Demo Video URL:} \url{https://youtu.be/bxx3bM5Pm6c}
  \end{flushright}
\end{abstract}



\maketitle

\section{Introduction}
\label{sec:intro}

In the last five years
much outstanding research has been devoted to showing how to 
exploit Bitcoin to execute \emph{smart contracts} ---
computer protocols which allow for exchanging cryptocurrency
according to complex pre-agreed rules
\cite{Andrychowicz14sp,Andrychowicz14formats,Andrychowicz14bw,Andrychowicz16cacm,Banasik16esorics,BZ17bw,Bentov14crypto,Kumaresan14ccs,Kumaresan15ccs,KumaresanB16ccs,KumaresanVV16ccs,Miller16zerocollateral}.
Despite the wide variety of use cases witnessed by these works,
no tool support has been provided yet to facilitate 
the development of Bitcoin contracts.
Today, this task requires to devise complex protocols which, 
besides using the standard cryptographic primitives,
can read and append transactions on the Bitcoin blockchain.
Creating a new protocol
requires a significant effort to establish its correctness and security:
this is an error-prone task, 
usually performed manually.
Crafting the transactions used by these protocols is burdensome as well,
since it requires to struggle with low-level, poorly documented
features of Bitcoin.

In this paper we consider BitML, 
a recent high-level language for smart contracts,
featuring a computationally sound embedding into Bitcoin~\cite{BZ18bitml},
and a sound and complete verification technique
of relevant trace properties~\cite{BZ19post}.
BitML can express many of the smart contracts in 
the literature~\cite{BCZ18isola,bitcoinsok},
and execute them by appending suitable transactions 
to the Bitcoin blockchain.
The computational soundness of the embedding guarantees that 
security properties at the level of the BitML semantics
are preserved at the level of Bitcoin transactions, even in the presence of adversaries.
Still, BitML lives in a theoretical limbo, as no tool support exists yet 
to develop contracts and deploy them on the Bitcoin blockchain.



  \paragraph{Contributions}

We develop a toolchain for writing and verifying BitML contracts,
and for deploying them on Bitcoin.
More specifically, our main contributions can be summarised as follows:
\begin{enumerate}

\item A BitML embedding in Racket~\cite{Flatt12cacm},
  which allows for programming BitML contracts within the DrRacket IDE.

\item A security analyzer which can check arbitrary LTL properties of BitML contracts.
  In particular, the analysis can decide \emph{liquidity}, 
  a landmark property of smart contracts  
  requiring that the funds within a contract do not remain frozen forever.

\item A compiler from BitML contracts to standard Bitcoin transactions.
  The computational soundness result in~\cite{BZ18bitml}
  ensures that attacks to compiled contracts
  are also observable at the BitML level.
  Therefore, the properties verified by our security analyzer
  also hold for compiled contracts.

\item A collection of BitML contracts,
  which we use as a benchmark to evaluate our toolchain.
  This collection contains some of the most complex contracts
  ever developed for Bitcoin,
  \eg financial services, auctions, timed commitments, lotteries, 
  and a variety of other gambling games.
  We use our benchmarks to discuss the expressiveness 
  and the limitations of Bitcoin as a smart contracts platform.

\end{enumerate}

\begin{figure}
  \resizebox{\columnwidth}{!}{
    \begin{tikzpicture}[>=triangle 45]
      \node[block] (a) {BitML \\ on DrRacket};
      \node[inputblock, below =0.7cm of a]   (ap){Properties + \\ Strategies};
      \node[inputblock, above =0.7cm of a]   (ac){Contract};

      \node[block, above right = 0.2cm and 2cm of a] (b) {Abstract BitML \\ semantics};
      \node[block, right =0.7cm of b]   (mc){Model \\ checker};
      \node[outputblock,right =1cm of mc]   (qr){Query \\ result};

      \node[block, below =2cm of b]   (c1){BitML to \\ Balzac};
      \node[block, right =0.7cm of c1]   (c2){Balzac to \\ Bitcoin};
      \node[outputblock,right =1cm of c2]   (tx){Bitcoin \\ transactions};

      \node at (4.2,3) {\textsc{Security Analyzer}};
      \draw [ultra thick, draw=black, fill=gray, opacity=0.05]
      (2.8,0.5) -- (2.8,2.8) -- (9.4,2.8) -- (9.4,0.5) -- cycle;

      \node at (3.5,-0.5) {\textsc{Compiler}};
      \draw [ultra thick, draw=black, fill=gray, opacity=0.05]
      (2.8,-2.9) -- (2.8,-0.7) -- (9.4,-0.7) -- (9.4,-2.9) -- cycle;

      \draw[arrows=->,line width=0.5pt] (ac.south) -| (a.north);
      \draw[arrows=->,line width=0.5pt] (ap.north) -| (a.south);

      \draw[arrows=->,line width=0.5pt] (a.east) -| (2.25,0) |- (b.west);
      \draw[arrows=->,line width=0.5pt] (a.east) -| (2.25,0) |- (c1.west);
      \draw[arrows=->,line width=0.5pt] (b.east) |- (mc.west);
      \draw[arrows=->,line width=0.5pt] (mc.east) |- (qr.west);

      \draw[arrows=->,line width=0.5pt] (c1.east) |- (c2.west);
      \draw[arrows=->,line width=0.5pt] (c2.east) |- (tx.west);
    \end{tikzpicture}
  }
    \vspace{-15pt}
  \caption{Toolchain architecture.}
  \label{fig:toolchain-architecture}

\end{figure}
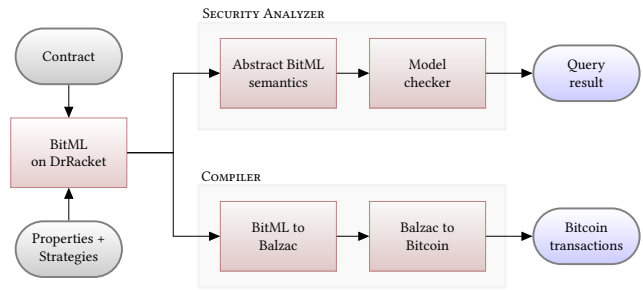

The architecture of our toolchain is displayed in~\Cref{fig:toolchain-architecture}.
The development workflow is the following:
\begin{inlinelist}[(a)]
\item write the BitML contract, and specify the required properties.
  Optionally, specify some constraints on the participants' strategies,
  \eg to partially define the behaviour of the honest participants;
\item verify that the contract satisfies the required properties 
  through the security analyzer;
\item compile the contract to Bitcoin transactions;
\item execute the contract, by appending these transactions to the Bitcoin blockchain
  according to the chosen strategy.
\end{inlinelist}
We remark that the last step can be performed on the Bitcoin main network, 
without requiring any extensions or customizations.
Our toolchain is open-source%
\footnote{\url{https://github.com/bitml-lang}},
as well as the contracts in our benchmark.
A tutorial is available online%
\footnote{\url{https://blockchain.unica.it/bitml}},
including references to our experiments on the Bitcoin testnet.

\section{Designing BitML contracts}
\label{sec:bitml}

BitML contracts allow two or more participants to exchange their bitcoins ($\BTC$)
according to a given logic.
A contract consists of two parts: 
a precondition, describing requirements that participants must fulfil to stipulate the contract,
and a process, which specifies the execution logic of the contract.
Here, rather than providing the syntax and semantics of BitML 
(see \cite{BZ18bitml} for a formalization),
we illustrate it through a simple but paradigmatic example, 
the \emph{mutual timed commitment} contract~\cite{Andrychowicz14sp}.
This contract involves two participants (named below \code{A} and \code{B})
each one choosing a secret and depositing a certain amount of cryptocurrency (say, $1 \BTC$).
The goal of the contract is to ensure that each participant will either learn the other participant’s secret, or otherwise receive the other participant’s deposit as a compensation.
The contract gives some time to the participants to reveal their secrets.
If a participant reveals her secret in time, then she can get her deposit back;
otherwise, after the time is up, the other participant can withdraw that deposit.

In our tool, we can specify this contract as follows:

\begin{center}
\scalebox{1.1}{
\begin{lstlisting}[language=bitml,classoffset=1,morekeywords={},classoffset=2,morekeywords={FundsA,Commit,Reveal},framexbottommargin=0pt,framextopmargin=0pt]
(participant "A" "029c...cced") ; A's public key
(participant "B" "022c...af30") ; B's public key

(contract
  (pre
    (deposit "A" 1 "1a34...6f38@0") ; tx output (1BTC)
    (secret  "A" a "628f...de71")   ; hash of A's secret
    (deposit "B" 1 "19e7...85ff@2") ; tx output (1BTC)
    (secret  "B" b "9d48...bb35"))  ; hash of B's secret

  (choice
    (reveal (a) (choice
                  (reveal (b) (split
		                (1 -> (withdraw "A"))
                                (1 -> (withdraw "B"))))
                  (after 100050 (withdraw "A"))))
    (after 100000 (withdraw "B")))
\end{lstlisting}
}
\end{center}

The first two lines create aliases for the participant names, specifying their public keys.
The contract preconditions are in the \code{pre} part:
each participant must specify the identifier of a transaction output, 
and the hash of the chosen secret.
The transaction output must be unspent, must contain the required $1 \BTC$, 
and must be redeemable using the participant’s private key.
The hash is used during the contract execution:
when the participant provides a value, claiming that it is the chosen secret,
the hash of this value is required to be equal to the one in the precondition. 

The contract logic is specified after the preconditions.
The top-level \code{choice} defines two alternative branches of the contract.
The first branch can only be taken if \code{A} reveals her secret (named \code{a});
when this happens, the contract continues with the innermost \code{choice}.
The second branch can only be taken after a timeout, 
specified as the block at height 100000,
and it allows \code{B} to redeem all the funds deposited within the contract (\ie, $2 \BTC$)
by executing \code{withdraw ``B''}.
So, to avoid losing her deposit, \code{A} is incentivized to reveal her secret in time.
Similarly, the innermost \code{choice} is used to incentivize \code{B} to reveal his secret
before the block at height 100050.
If \code{B} reveals, then the \code{split} subcontract is executed:
this divides the balance of the contract in two parts of $1 \BTC$ each, 
allowing the participants to withdraw their deposits back.

The language is defined exploiting the Racket macro system,
which is used to rewrite BitML syntactic constructs to Racket code. 
This approach benefits from the Racket language ecosystem,
and allows us to write BitML contracts in the DrRacket IDE.
Indeed, our toolchain integrates within the DrRacket IDE
the contract editor, the security analyzer and the BitML compiler.
The implementation of BitML in Racket extends the idealized version of BitML in~\cite{BZ18bitml} 
to make the language usable in practice.
For instance, it introduces special deposits of type \code{fee}, 
which are automatically spread over all the transactions obtained by the compiler.
We also implement static checks for a number
of errors that could prevent the correct execution of contracts,
\eg committing secrets with the same hash, 
double spending a transaction output, \etc

\section{Verifying BitML contracts}
\label{sec:verification}

The tool verifies various forms of \emph{liquidity},
requiring that no funds (or funds up-to a certain amount) are frozen forever within a contract%
\footnote{A paradigmatic case of non-liquid contract was 
 the Ethereum Parity Wallet~\cite{parity17nov}.
 An attacker managed to kill a library called by the wallet, 
irreversibly freezing ${\sim}160M \$$.}.
Further, the tool can verify arbitrary LTL formulae, where state predicates
can specify, \eg, the funds owned by participants, the provided authorizations, and the revealed secrets.
By default, the tool verifies the required property against \emph{all} possible behaviours of each participant: for instance, if a contract contains \code{reveal a},
the verifier considers both the case where the secret is revealed 
and the one where it is not.
Authorizations are handled similarly, by considering both cases.
However, in most cases, a participant wishes to verify a contract with respect to 
a given behaviour for herself,
making no assumptions on the other participants’ behaviour
(unless some other participants are considered trusted, in which case it would make sense to fix a behaviour also for them).
For instance, a participant \code{A} may want to give her authorization
to perform a given branch only after participant \code{B} has revealed his secret.
The tool allows for constraining the behaviour of participants,
specifying the conditions upon which secrets are revealed
and authorizations are provided.
Actions which can be performed by everyone,
like \code{withdraw} and \code{split}, cannot be constrained.

For instance, we can verify that the mutual timed commitment contract is liquid
whatever strategies are chosen by participants.
The query \code{check-liquid} correctly answers true, since:
\begin{inlinelist}
  \item if \code{A} does not reveal, then anyone (after the block at height 100000)
    can perform \code{withdraw ``B''}, which transfers the whole contract balance to \code{B};
  \item if \code{A} reveals but \code{B} does not reveal, then anyone (after the block at height 100050)
    can perform \code{withdraw ``A''}, which transfers the whole contract balance to \code{A};
  \item if both \code{A} and \code{B} reveal, then anyone can perform \code{split},
    which transfers the balance in equal parts to \code{A} and \code{B}.
\end{inlinelist}

Note that if we remove the \code{after} branch at line~\lineno{16}, 
the contract is no longer liquid.
However, it becomes liquid when \code{A}'s strategy is to reveal the secret. 
We can verify that this holds through the query
\code{check-liquid (strategy "A" (do-reveal a))}.
Liquidity is lost again 
if \code{A} chooses to reveal only after \code{B} has revealed,
\ie when her strategy is
\code{"A" (do-reveal a) if ("B" (do-reveal b))}.

Besides liquidity, we can check specific LTL properties of contracts 
through the command \code{check-query}.
\Eg, in the mutual timed commitment we can verify that,
after \code{A} reveals, 
she will eventually get back at least her $1 \BTC$ deposit.
In LTL, this property is formalised as the following formula, where $10^8$ satoshi = 1\BTC:
\begin{center}%
\scalebox{0.95}{%
\code{[](a revealed => <>A has-deposit>= 100000000 satoshi)}}
\end{center}
We also verify that if \code{A} reveals the secret, 
then eventually either \code{B} reveals, or \code{A} will get \code{B}'s deposit,
too.
The LTL query is the following:
\begin{flushleft}
\scalebox{0.95}{\;\code{[](a revealed =>}}
\scalebox{0.95}{\;\code{<>(b revealed \codebackslash/ A has-deposit>= 200000000 satoshi))}}
\end{flushleft}


Our verification technique is based on model-checking the state space of BitML contracts.
Since this state space is infinite, before running the model-checker
we reduce it to a finite-state one, by exploiting the abstraction in~\cite{BZ19post}.
This abstraction resolves the three sources of infiniteness
of the concrete semantics of BitML:
the passing of time,
the advertisement/stipulation of contracts,
and the off-contract bitcoin transfers.
To obtain a finite-state system, the abstraction:
\begin{inlinelist}
\item quotients time in a finite number of time intervals, 
\item disables the advertisement of new contracts, and 
\item limits the off-contract operations to those
for transferring funds to contracts and for destroying them.
\end{inlinelist}
This abstraction is shown in~\cite{BZ19post} to enjoy 
a strict correspondence with the concrete BitML semantics:
namely, each concrete step of the contract under analysis
is mimicked by an abstract step, and vice versa.

Our tool implements the abstract BitML semantics in Maude,
a model-checking framework based on rewriting logic~\cite{Maude01}.
Maude is particularly convenient for this purpose:
we use its equational logic to express 
structural equivalence between BitML terms, 
and its conditional rewriting rules to encode the abstract semantics of BitML.
In this way, we naturally obtain an executable abstract semantics of BitML.
Once a BitML contract in translated in Maude, 
we use the Maude LTL model-checker~\cite{Eker02maude} 
to verify the required security properties, under the strategies specified by the user. 
The various forms of liquidity are also translated to corresponding LTL formulae.
The computational soundness of the BitML compiler
guarantees that the properties verified by the model checker are preserved when executing 
the contract on Bitcoin.

\section{Compiling BitML to Bitcoin}

Our compiler operates in two phases:
first, it translates BitML contracts into Balzac%
\footnote{\url{https://github.com/balzac-lang/balzac}}, 
an abstraction layer over Bitcoin transactions based 
on the formal model of~\cite{bitcointxm};
then, it translates Balzac transactions
into standard Bitcoin transactions.
The compiler from BitML to Balzac implements the algorithm in~\cite{BZ18bitml},
extending it with transaction fees.
In particular, the compiler guarantees that each transaction contains enough
fees to be publishable in the blockchain.
The compiler from Balzac to Bitcoin produces
\emph{standard} Bitcoin transactions~\cite{standard-tx}:
this is crucial since non-standard ones are discarded by the Bitcoin network. 
To this aim, Balzac produces standard output scripts 
of the form ``Pay to Public Key Hash'' (P2PKH) or ``Pay to Script Hash'' (P2SH).
P2PKH is used for encoding signature verification
(\eg, to redeem the deposit obtained by a \code{withdraw}),
while P2SH is used for complex redeeming conditions
(\eg, to check that the revealed secret matches the committed hash).
Since Bitcoin requires that all the values pushed by standard scripts 
fit within 520 bytes, our compiler checks that this constraint is satisfied
for each generated script.
Balzac outputs serialized raw transactions, 
which can be directly broadcast to the Bitcoin network.

\section{Evaluation}
\label{sec:evaluation}

To evaluate our toolchain, we use a benchmark of representative use cases,
including financial contracts~\cite{Thompson18isola,Biryukov17wtsc}, 
auctions, lotteries~\cite{Andrychowicz16cacm,Miller16zerocollateral} 
and gambling games%
\footnote{\url{https://github.com/bitml-lang/bitml-compiler/tree/master/examples/benchmarks}}.
For each contract in the benchmark, 
we display in \Cref{fig:evaluation:benchmarks}
the number $N$ of involved participants,
the number $T$ of transactions obtained by the compiler,
and the verification time $V$ for checking liquidity%
\footnote{For uniformity, in the performance evaluation we focus on liquidity
We carry out our experiments on a PC with
a Intel Core i7-7800X CPU @ 3.50GHz, and 64GB of RAM.}.
The participants' strategies are constrained only as needed to ensure liquidity:
in most cases, we do not put any constraints at all.
For the contracts which involve predicates on secrets (\eg, all the lotteries), 
in principle one would need to check liquidity against all the possible choices of secrets.
To make verification feasible, 
since each contract only checks a finite set of predicates,
we partition the infinite choices of secrets into a finite set of regions,
and sample one choice from each region.
In this way, the liquidity check is performed a finite number of times,
ensuring that the verifier explores every reachable state of the contract.
For instance, in the 4-players lottery we explore $3^4$ regions,
which explains the 67 hours needed to verify its liquidity.%
\footnote{Another feature which significantly affects the verification time 
is the fact that we are considering \emph{all} the possible strategies
of \emph{all} the participants.}




\begin{table}[t!]
  \centering
  \small
  \begin{tabular}{|c|c|c|c|}
    \hline
    \textbf{Contract} & \textit{\textbf{N}} & \textit{\textbf{T}} & \textit{\textbf{V}} \\
    \hline
    Mutual timed commitment & 2 & 15 & 83ms \\
    Mutual timed commitment & 3 & 34 & 103ms \\
    Mutual timed commitment & 4 & 75 & 454ms \\
    Mutual timed commitment & 5 & 164 & 13s \\
    Escrow (early fees) & 3 & 12 & 8s \\    
    Escrow (late fees) & 3 & 11 & 3.4s \\ 
    Zero Coupon Bond & 3 & 8 & 86ms \\
    Coupon Bond & 3 & 18 & 1.3s \\
    Future$(C)$ & 3 & 5 + $\mathit{T}_C$ & 80ms + $V_C$ \\
    Option$(C,D)$ & 3 & 14 + $T_{C} + T_{D}$ & 90ms + $V_C + V_D$ \\
    Lottery ($O(N^2)$ collateral) & 2 & 15 & 427ms \\
    Lottery ($0$ collateral) & 2 & 8 & 142ms \\
    Lottery ($0$ collateral) & 4 & 587 & 67h \\
    Rock-Paper-Scissors & 2 & 23 & 781ms \\
    Morra game & 2 & 40 & 674ms \\
    Shell game & 2 & 23 & 27s \\
    Auction (2 turns) & 2 & 42 & 3.3s \\
    \hline
  \end{tabular}
  \caption{Benchmarks for the BitML toolchain.}
  \vspace{-10pt}
  \label{fig:evaluation:benchmarks}
\end{table}

The only work against which we can compare the performance of 
our tool is~\cite{Andrychowicz14formats}, 
which models Bitcoin contracts in Uppaal,
a model-checking framework based on Timed Automata.
The most complex contract modelled in~\cite{Andrychowicz14formats}
is the mutual timed commitment with 2 participants:
this requires \mbox{$\sim 30$s} to be verified in Uppaal,
while our tool verifies the same property in \mbox{$< 100$ms}.
This speedup is due to the higher abstraction level of BitML
over~\cite{Andrychowicz14formats}, 
which operates at the (lower) level of Bitcoin transactions.

One of the main difficulties that we have encountered 
in developing contracts is that some complex BitML specifications
can not be compiled to Bitcoin,
because Bitcoin has a 520-byte limit on the size of each value pushed to the 
evaluation stack~\cite{btc520bytes}.
In some cases, we managed to massage the BitML contract
so to make its compilation respect the 520-byte constraint.
For instance, a common pattern  that easily violates the 520-byte constraint
is the following:
\begin{center}
\scalebox{1.18}{
\begin{lstlisting}[language=bitml,classoffset=1,morekeywords={},classoffset=2,morekeywords={FundsA,Commit,Reveal},framexbottommargin=0pt,framextopmargin=0pt]
(choice (revealif (b) (pred (p0)) (C0))
        (revealif (b) (pred (p1)) (C1))
        (after T      (C2)))
\end{lstlisting}
}
\end{center}

The \code{choice} is compiled
into a transaction whose redeem script encodes
the disjunction of \emph{three} logical conditions,
corresponding to the three branches of the \code{choice}.
Depending on the predicates \code{p0} and \code{p1},
and on the number of participants in the contract,
this script may violate the 520-byte constraint.
A workaround is to rewrite the pattern above into the following one:
\vspace{-10pt}
\begin{center}
\scalebox{1.18}{
\begin{lstlisting}[language=bitml,classoffset=1,morekeywords={},classoffset=2,morekeywords={FundsA,Commit,Reveal},framexbottommargin=0pt,framextopmargin=0pt]
(choice (revealif (b) (pred (p0)) (C0))
        (after T (tau (choice
                      (revealif (b) (pred (p1)) (C1))
                      (after T1     (C2))))))
\end{lstlisting}
}
\end{center}

In this case the compilation includes two transactions, 
corresponding to the two \code{choice}s.
The scripts of these transactions encode the disjunction of \emph{two} 
logical conditions, corresponding to the two branches of the \code{choice}s.
Using this workaround we have managed to compile the 4-players lottery
into standard transactions, 
at the price of increasing the number of transactions 
(587 for the standard version \emph{vs.} 138 for the nonstandard one).
Similar techniques (\eg simplification of predicates) allowed us to compile 
all the contracts in~\Cref{fig:evaluation:benchmarks}
into standard Bitcoin transactions.

In general, the 520-byte constraint intrinsically 
limits the expressiveness of Bitcoin contracts:
for instance, since public keys are 33 bytes long, 
a contract which needs to simultaneously verify 15 signatures
can not be implemented using standard transactions.

\section{Conclusions}



Although our benchmarks witness a rich variety of contracts expressible in BitML,
there is room for improvement. 
BitML is not \emph{Bitcoin-complete}, 
\ie some contracts executable in Bitcoin are not expressible in BitML.
The main sources of this incompleteness are three:
\begin{inlinelist}
\item all the transactions obtained by the compiler must be signed  \emph{before} stipulation
by all the involved participants (only the signatures for authorizations can be provided at run-time);
\item all transaction fields must be taken into account when computing signatures,
while partial signatures (\eg those obtained through
\code{SIGHASH\_ANYONECANPAY} and \code{SIGHASH\_SINGLE}) are not used;
\item off-chain interactions are limited to revealing secrets
and providing authorizations.
\end{inlinelist}
The first constraint is required to ensure that honest participants 
can always perform, at the Bitcoin level, the moves enabled in the 
corresponding BitML contract, regardless of the behaviour of the others.
In this respect, BitML follows the standard assumption
that participants 
are \emph{non-cooperative}, \ie at any moment after stipulation they can stop
interacting
(unlike TypeCoin~\cite{Crary15pldi}, which assumes cooperation,
allowing dishonest participants to make a contract deadlock).
Yet, cooperation can be incentivized,
by punishing misbehaviour with penalties, 
like \eg in the timed commitment of~\Cref{sec:bitml}.
As a consequence of the design choices above, 
contracts with a dynamically-defined set of players
(\eg, crowdfunding), 
or an unbounded number of iterations (\eg, micro-payment channels), 
are not expressible in BitML. 

The limitations of BitML (and of Bitcoin) could be overcome in various ways.
For instance, using Bitcoin ``as-is'', 
it would be possible to relax constraint (iii) above, 
so to allow \eg zero-knowledge off-chain protocols.
This would enable to extend BitML with primitives to 
express \emph{contingent payments} contracts,
where participants trade solutions of a class of NP problems~\cite{Banasik16esorics,bitcoin-zkcp}.
Similarly, by relaxing constraint (i), 
we could extend BitML to enable dynamic stipulation of subcontracts,
requiring that all the involved participants provide their signatures at run-time.
This would allow to model \eg micro-payment channels in BitML.
Together with the use of \code{SIGHASH\_ANYONECANPAY}
(relaxing constraint (ii)),
this would also allow for modelling crowdfunding contracts.
As before, this extension could be implemented without modifying Bitcoin.

Other extensions of BitML would require extensions of Bitcoin.
For instance, \emph{covenants}~\cite{Moser16bw,Oconnor17bw} 
would allow for implementing arbitrary finite-state machines.
%
Controlled \emph{input malleability} would allow to efficiently 
implement tournaments in multi-player gambling games, 
like \eg lotteries~\cite{BZ17bw}.
This can also be achieved through a new opcode 
that checks if the redeeming transaction belongs to a 
given set~\cite{Miller16zerocollateral}.
Contingent payments without zero-knowledge proofs
can be achieved by exploiting a new opcode that checks 
the validity of key pairs~\cite{delgado2017fair}.
A new opcode which checks signatures for arbitrary messages
would allow for expressing general fair multiparty computations~\cite{Kumaresan15ccs}.
Further, fair and robust multiparty computations can be achieved 
using more complex transactions~\cite{Kiayias16eurocrypt}.
A more radical approach would be to replace the Bitcoin scripting language
with a more expressive one, like \eg Simplicity~\cite{Oconnor17plas}.

Compared with the tools for analysing Ethereum contracts
\cite{Luu16ccs,Tsankov18securify,Grishchenko18post,Hildenbrandt18csf,Park18sigsoft,Grishchenko18cav,Bhargavan16solidether,Sergey18scilla},
whose precision is subject to the limitations
derived by the Turing-completeness of the underlying languages,
our toolchain features a sound and complete verification technique.


\begin{acks}
Work partially supported
by MIUR PON 2018 ``Distributed Ledgers for Secure Open Communities'' ARS01\_00587;
by R.A.S.
projects ``Sardcoin'' and ``Smart collaborative engineering'';
by EPSRC projects EP/K034413/1,
EP/K011715/1, EP/L00058X/1, EP/N027833/1 and EP/N028201.
Stefano Lande is supported by 
P.O.R. F.S.E. 2014-2020.
\end{acks}

\bibliographystyle{ACM-Reference-Format}
\bibliography{main}



\end{document}